\documentclass[final,5p,times,twocolumn]{elsarticle}

\usepackage{graphicx}

\usepackage{amssymb}

\biboptions{compress}

\journal{Physica C: Superconductivity and its Applications}

\begin{document}

\begin{frontmatter}


\title{Cu NMR study of a Pb-doped Bi:2201 single crystal.}


\author [label1]{O.~N.~Bychina}
\author [label2]{A.~N.~Maljuk}
\author [label2,label3]{S.~Wurmehl}
\author [label2]{B.~B\"{u}chner}
\author [label4]{O.~M.~Vyaselev\corref{cor1}}
\ead{vyasel@issp.ac.ru}
\address[label1]{Moscow Institute of Physics and Technology, 141700 Dolgoprudny, Russia}
\address[label2]{Leibniz-Institut f\"{u}r Festk\"{o}rper- und Werkstoffforschung Dresden, Helmholtzstra{\ss}e
20, D-01069 Dresden, Germany}
\address[label3]{Institut f\"{u}r Festk\"{o}rperphysik, Technische Universit\"{a}t Dresden, D-01171 Dresden,
Germany}
\address[label4]{Institute of Solid State Physics, Russian Academy of Sciences, 142432 Chernogolovka, Russia}

\cortext[cor1]{Corresponding author. Tel.: +7 49652 28374; Fax: +7 496 524 9701.}

\begin{abstract}
A $^{63}$Cu NMR study of Pb-doped Bi:2201 system, Bi$_{1.6}$Pb$_{0.4}$Sr$_{2.05}$CuO$_y$, is
presented. Temperature dependencies of the NMR peak shift and the nuclear spin-lattice relaxation
rate revealed the pseudogap that opens at $T^{\ast}=20-60\,$K, way above the $T_c\simeq9\,$K
measured for the orientation ($H\parallel c$) and value (7\,T) of the NMR experiment field. The
noticeable discrepancy between $T_c$ and $T^{\ast}$ and the behavior of Cu SLR at $T>T^{\ast}$
imply the underdoped state of the studied system. The magnetic field has a relatively weak effect
on the superconductivity in the studied system, as evidenced from small (7-8\,K) shift of the
zero-field $T_c=16\pm 1\,$K under the applied 7\,T field. This fact suggests a high value of the
upper critical field, unusual for a compound with such low $T_c$.

\end{abstract}


\end{frontmatter}


\section{Introduction}

Since the discovery of High-temperature superconductivity (HTS) in 1986, an impressive advance of
material science during the subsequent decade has resulted in the enhancement of the
superconducting transition critical temperature ($T_c$) up to nearly 140\,K \cite{SchillingNat93}.
However, the extensive theoretical and experimental attempts to find a comprehensive excuse for the
HTS phenomenon have apparently been less successful, though many issues have been undoubtedly
resolved. Among them, for instance, the CuO$_2$ plane as the structural sub-unit responsible for
both the conductivity and superconductivity. A fingerprint of all HTS's is also the so-called
`pseudogap' -- the phenomenon unusual to common (low-temperature BCS-type) superconductors,
revealed in the decay of the density of states (DOS) at temperatures appreciably higher than $T_c$.
Initially, the pseudogap has been found by NMR \cite{HammelPRL89,TakigawaPRB91, WarrenPRL89} and
inelastic neutron scattering \cite{RosMig92} techniques; in both cases it manifests itself as a
drop of the electronic spin excitations at a temperature way above $T_c$. Later on, the effect has
been verified by tunnel experiments \cite{RennerPRL98} that tied the abnormal transport and spin
properties of HTS's above $T_c$.

Understanding the relationship between the normal-state and superconducting properties of the
copper oxides, in general, and of the nature of the pseudogap, in particular, is believed to be a
key issue in unlocking the mystery of HTS'c \cite{VarmaNat10}. The origin of the pseudogap has been
widely disputed over the past decades. Some theories attributed it to superconducting fluctuations
\cite{EmeryNat95}, others employed a two-gap scenario where the pseudogap is associated with
another order of non-superconducting origin, that coexists and competes with the superconductivity.
Scrutinized by various experimental techniques including NMR \cite{VyaselPRL99}, tunnel
spectroscopy \cite{KrasnovPRL00, KrasnovPRB02, KrasnovPRB12}, polarized elastic neutron diffraction
\cite{FauqPRL06}, and angle-resolved photoemission spectroscopy (ARPES) \cite{KondoPRL07}, this
dualism has been seemingly removed in favour of the non-superconducting nature of the pseudogap.
Recent X-ray scattering experiments revealed the formation of temperature dependent incommensurate
charge-density wave (CDW) fluctuations, supporting a theory in which the superconducting and
charge-density wave orders exhibit angular fluctuations in a six-dimensional space
\cite{SachdevSci14}.

In order to add more experimental facts that might clarify the issue of the pseudogap phenomenon,
we performed an NMR study of Pb-doped Bi:2201 (Bi$_2$Sr$_2$CuO$_6$) compound,
Bi$_{1.6}$Pb$_{0.4}$Sr$_{2.05}$CuO$_y$ ($y\approx6$). The choice of this particular system was
motivated by several reasons. First of all, the stoichiometric parent compound is either
non-superconducting or exhibits a very low $T_c<4\,$K; moreover, its crystal structure is
characterized by substantial modulation in BiO plane \cite{StrucBi2201PRB93}. Substitution of La
for Sr gives rise to superconductivity with $T_c$ up to 36\,K but does not remove the modulation,
while substitution of Pb for Bi makes the system both superconducting and modulation-free. The
latter makes it more suitable for ARPES studies since the corresponding spectra become easier to
interpret \cite{KondoPRL07}, which could be useful for further investigations. Next, the
replacement of trivalent Bi by bivalent Pb acts, in the sense of the charge-doping, in the way
opposite to the replacement of Sr$^{2+}$ by La$^{3+}$. The La-doped Bi:2201 compound was
extensively studied by NMR recently \cite{ZhengPRL10} while the NMR data for its Pb-doped relative
has been missing. Finally, relatively low transition temperature of Pb-doped Bi:2201
($T_{c0}\lesssim20$\,K) assumes low upper critical field $H_{c2}$, that gives a perspective to
suppress superconductivity with a reasonable magnetic field and track the normal-state properties
(including the pseudogap) down to temperatures far below $T_{c0}$.

In this paper we report on the first $^{63}$Cu NMR results obtained for
Bi$_{1.6}$Pb$_{0.4}$Sr$_{2.05}$CuO$_y$. We show that temperature dependencies of both the Knight
shift and the nuclear spin-lattice relaxation rate measured in 7\,T field applied along the
\textit{c}-axis of the crystal, demonstrate a pseudogap behaviour below $\simeq50\,$K, far above
the superconducting transition temperature of 8-9\,K recorded for this field value and sample
orientation. The substantial disparity between the pseudogap and the superconducting gap implies
the underdoped state of the compound in study \cite{KrasnovPRB12,ZhengPRL05}.

\section{Experimental}


Large size and high quality Bi$_{1.6}$Pb$_{0.4}$Sr$_{2.05}$CuO$_y$ single crystals have been grown
by the crucible-free floating zone method similar to \cite{KumagaiPhC97, MaljukPhC01}. The
four-mirror type image furnace (CSI, Japan), equipped with 4$\times$300\,W halogen lamps, was
employed for the crystal growth. The growth was performed in Ar/O$_2$ = 90\%/10\% gas mixture. The
pulling rate was 0.8-1.0\,mm/h. Single crystals were cleaved mechanically from the as-grown ingot.
The largest samples have dimensions up to 15$\times$4$\times$3\,mm$^3$ with a clear cleavage plane
of the (010) type. Preliminary neutron diffraction experiments performed at E4 beam-line (HZB,
Berlin) showed that neutron rocking curves of the samples that are 2-3\,mm thick along the
\textit{c}-axis, have shoulders around some Bragg peaks indicating the presence of small grains whose
orientation slightly differs from the bulk of the crystal. The mosaicity of thick Pb-doped Bi-2201
crystals can thus be estimated within 2-3$^\circ$. In contrast, the crystals with the thickness 0.3-0.5\,mm
demonstrated much lower mosaicity of about 0.5-0.7$^\circ$. In all cases, no traces of
incommensurate superstructure was seen in reciprocal space survey. Also, preliminary ARPES
measurements (BESSY, Berlin) gave perfect ARPES spectra being free from any modulation
complications. The $T_c$ of as-grown samples was around 4\,K while the Ar-annealed crystals
demonstrated $T_c\approx17$\,K. Powder X-ray diffraction on crushed samples confirmed that all
as-grown samples are impurity-free and possess the orthorhombic structure.

One single crystal with dimensions $\approx3\times5\,$mm$^2$ in the (\textit{ab}) plane and
$\approx1\,$mm along \textit{c}-axis was used for the NMR measurements. The sample placed inside the
NMR coil, was rigidly mounted on the probehead with its \textit{c}-axis aligned with the
longitudinal axis of the measurement insert, which coincides with the direction of the external
static magnetic field. Variation of the sample temperature was provided by the Oxford CF-1200 flow
cryostat. The transition temperature of the sample in study was measured by monitoring the NMR tank
circuit resonance frequency, $\nu_t$ (ca.\,80\,MHz). As the sample goes superconducting, the pickup
coil inductance, $L\propto(\chi+1)$ diminishes following the decrease of the ac susceptibility of the
sample, $\chi$ (due to the buildup of the shielding superconducting current), and
$\nu_t\propto1/\sqrt{L}$ shifts upwards. The measurements were made in the field $\mu_0H=7\,$T
(with the CF-1200 cryostat inserted into the room-temperature bore of the magnet) and at zero field
(the CF-1200 placed outside the magnet).

The NMR data was measured using a Bruker MSL-300 spectrometer in the external field $\mu_0H=7\,$T.
Standard spin-echo, saturation-recovery and inversion-recovery techniques with characteristic
$\pi/2$-pulse length of 1.5\,$\mu$s, were used to collect the spectra and measure the spin-lattice
relaxation (SLR) times. The position of broad ($\simeq0.75\,$MHz) NMR peaks was determined from
Gaussian fits to point-by-point-collected data. The SLR times, $T_1$, were extracted by fitting the
spin-echo area transients, $A(\tau)$, with the known formula \cite{NarathPR67}
$A(\tau)\propto0.9\textrm{e}^{6\tau/T_1}+0.1\textrm{e}^{\tau/T_1}$ for the central
$1/2\leftrightarrow -1/2$ magnetic transition of spin-3/2 Cu nucleus, where $\tau$ is the delay
after the saturation comb or the inversion pulse.


\section{Results and Discussion}

\begin{figure}[h]
\includegraphics[scale=0.3,angle=0]{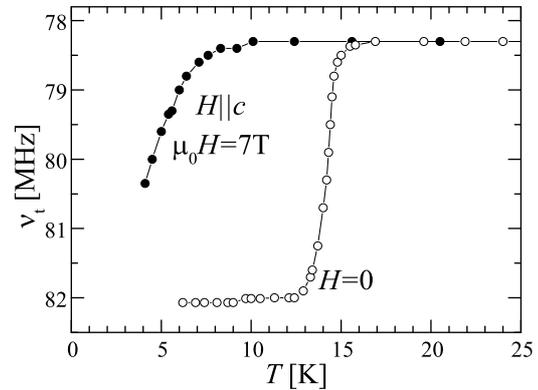} \caption{\label{Fig1} Superconducting
transition of the Bi$_{1.6}$Pb$_{0.4}$Sr$_{2.05}$CuO$_y$ single crystal in zero field (open
circles) and in field $\mu_0H=7\,$T (filled circles) aligned parallel to its \textit{c}-axis,
viewed as the shift of the NMR tank circuit resonance frequency.}
\end{figure}

Figure~\ref{Fig1} shows the NMR tank circuit resonance frequency, $\nu_t$, plotted against
temperature, for the sample outside the magnet ($H=0$) and in the magnetic field $\mu_0H=7\,$T
aligned with the \textit{c}-axis of the crystal. At zero field, with decreasing temperature, $\nu_t$ begins to diminish below
16-17\,K indicating the superconducting transition. This number coincides, to within 1\,K, with
preliminarily taken dc-susceptibility data measured in the zero-field-cooled mode, and with the
100\,kHz ac-susceptibility results. In the external field of 7\,T, the $T_c$ is shifted down to
8-9\,K.

\begin{figure}[bh]
\includegraphics[scale=0.3,angle=0]{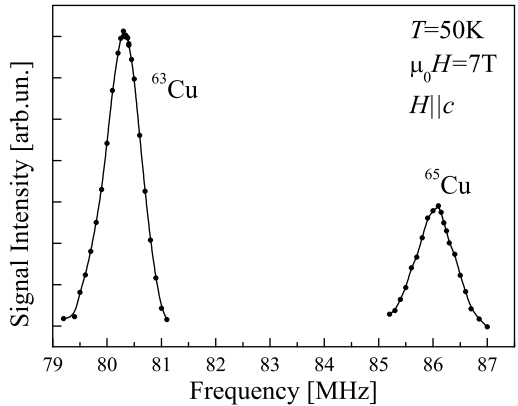} \caption{\label{Fig2} Cu NMR spectrum in
Bi$_{1.6}$Pb$_{0.4}$Sr$_{2.05}$CuO$_y$ for the $H\parallel c$ orientation, $\mu_0H=7\,$T,
$T=50\,$K.}
\end{figure}

The full NMR spectrum for the central ($1/2\leftrightarrow -1/2$) transition of Cu including its
$^{63}$Cu and $^{65}$Cu isotopes for the $H\parallel c$ orientation, measured at $\mu_0H=7\,$T,
$T=50\,$K, is shown in Figure~\ref{Fig2}. The full width at half maximum (FWHM) for $^{63}$Cu
resonance peak is $\sim750\,$kHz, which is nearly 40 times bigger than that in optimally-doped
YBa$_2$Cu$_3$O$_7$ \cite{MartinPRB94}. Apparently, the peak is broadened by the scatter of the
quadrupolar and/or hyperfine (orbital) shifts of the resonance frequency caused by structural
imperfections. For example, in La-doped Bi:2201 where the modulation in the BiO plane is present,
the FWHM of $^{63}$Cu peak is 1-2\,MHz \cite{ZhengPRL05}, whereas in Tl:2201 which has some
structural disorder in TlO planes \cite{VyaselPhC95}, the $^{63}$Cu peak is $\sim500\,$kHz wide
\cite{VyaselPdF96}.

\begin{figure}[t]
\center\includegraphics[scale=0.3,angle=0]{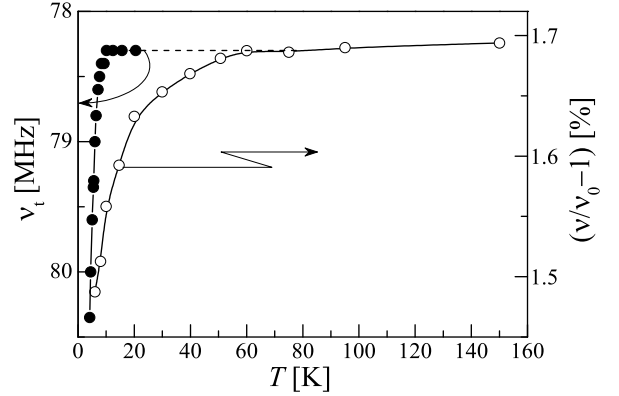} \caption{\label{Fig3} Open circles, right
axis: Temperature dependence of the total $^{63}$Cu NMR line shift, $\nu/\nu_0-1$, measured at
$\mu_0H=7\,$T, $H\parallel c$. Filled circles, left axis: $T$-dependence of the NMR tank circuit
resonance frequency for the same crystal orientation and value of the magnetic field (replotted
from Fig.~\ref{Fig1}).}
\end{figure}

The temperature dependence of $^{63}$Cu NMR line shift in Pb-doped Bi:2201 is shown in
Figure~\ref{Fig3}. The shift is calculated as $\nu/\nu_0-1$, where $\nu$ is the measured frequency
of the peak and $\nu_0 = ^{63}\gamma H/2\pi$ ($^{63}\gamma$ is the gyromagnetic ratio for
$^{63}$Cu). For the sake of visual comparison we added the data from Fig.~\ref{Fig1} for the same
crystal orientation and value of the magnetic field, depicting the change of the NMR tank circuit
resonance frequency at $T_c$ due to the shielding effect of superconducting currents. Quite
obviously, the position of $^{63}$Cu peak starts to decrease upon cooling at $T=55\pm5\,K$,
appreciably higher than $T_c$ for this field.

In a solid, the total frequency shift of an NMR peak from the position $\nu_0=\gamma H/2\pi$
resulted from the interactions of the nuclear spin $I$ with surrounding electrons, is contributed
from quadrupolar (for $I>1/2$) and magnetic shifts. In our particular case of the central
transition and $H\parallel c$ geometry, the quadrupolar shift is zero since the field is aligned
with the symmetry axis of the electron gradient field at Cu site. The magnetic (hyperfine) shifts
\cite{MacLaughlin76} include the diamagnetic contribution from the core electrons of the atom, the
orbital shift related to the Van-Vleck susceptibility, and the Knight shift resulted from contact
interaction with conduction electron spins, $K_s\propto\chi_s$, where $\chi_s$ is the conduction
electron static spin susceptibility. The former two contributions are generally
temperature-independent while $K_s$ follows the $T$-dependence of $\chi_s$ coupled to the density
of states (DOS) at the Fermi level. Therefore, the decrease observed in the temperature dependence
of the total shift shown in Fig.~\ref{Fig3} is associated with $K_s(T)$ and denotes the decrease in
DOS, hence opening of a spin gap. The temperature where it occurs is much higher than $T_c$ which
implies the pseudogap.

\begin{figure}[h]
\center\includegraphics[scale=0.3,angle=0]{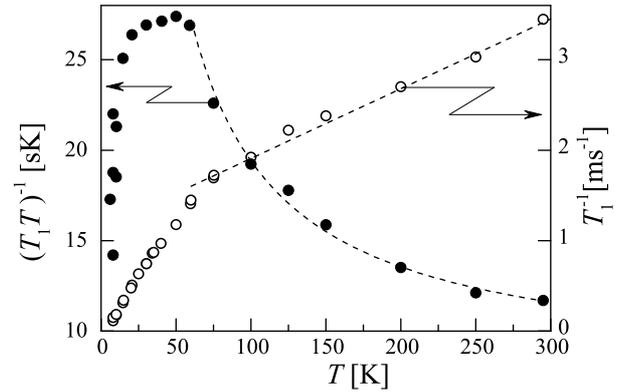} \caption{\label{Fig4} The data for $^{63}$Cu
nuclear SLR temperature dependence, $\mu_0H=7\,$T, $H\parallel c$. Open circles, right axis: The
SLR rate, $T_1^{-1}$. Filled circles, left axis: The SLR rate normalized to temperature,
$(T_1T)^{-1}$. Dashed lines are guides to the eye.}
\end{figure}

The data for $^{63}$Cu nuclear SLR temperature dependence measured in
Bi$_{1.6}$Pb$_{0.4}$Sr$_{2.05}$CuO$_y$ single crystal ($\mu_0H=7\,$T, $H\parallel c$) are depicted
in Figure~\ref{Fig4}. The $T$-dependence of $T_1^{-1}$ is nearly linear from room temperature down
to 75\,K as guided by the dashed line, and decreases upon further cooling more rapidly. As for the
$(T_1T)^{-1}$ product, it increases in a hyperbolic manner upon cooling down to 60\,K and has a
broad maximum at $T=40\pm20\,$K. The temperature where $(T_1T)^{-1}$ starts to decrease can be argued
between 50 and 30\,K, in both cases well above the $T_c\simeq9\,$K of the sample in 7\,T field
parallel to \textit{c} axis (see Fig.~\ref{Fig1}).

The temperature behaviour of Cu SLR shown in Fig.~\ref{Fig4} is typical for HTC cuprates
\cite{HammelPRL89,TakigawaPRB91,ZhengPRL05,KitaokaPRB98}. In a nonmagnetic metal the nuclear SLR is
governed by spin fluctuations of conduction electrons, generally expressed as $(T_1T)^{-1}\propto
\lim\limits_{\omega\rightarrow0}\sum\limits_{q}|A(q)|^2\chi^{ \prime\prime}(q,\omega)/\omega$,
where $A(q)$ is the hyperfine coupling between the electronic spin and the nuclear spin and $\chi^{
\prime\prime}(q,\omega)$ the imaginary part of the wave-vector- $(q)$ and frequency- $(\omega)$
dependent dynamic electronic spin susceptibility \cite{HammelPRL89}. In common metals where $\chi^{
\prime\prime}$ is basically $(q)$-independent,  the SLR and the Knight shift are related by the
well-known Korringa formula $(K_s^2T_1T)^{-1}\equiv\pi h k_B \gamma^2/\mu_B$ \cite{Korringa50}
which assumes the $T$-independence of the $(T_1T)^{-1}$ product (provided that $K_s$ is constant)
in the normal state. However, in HTS cuprates the Korringa relation is violated because $\chi^{
\prime\prime}$ is peaked at the antiferromagnetic wave vector $Q=(\pi,\pi)$ and $(T_1T)^{-1}$
reproduces the $(T+\Theta)^{-1}$-like behavior of $\chi^{ \prime\prime}(Q)$, caused by the
development of antiferromagnetic correlations \cite{ZhengPRL05}. At some temperature $T^{\ast}>T_c$
the $(T_1T)^{-1}$ product has a broad maximum and decreases upon further cooling due to the
pseudogap.

Such behavior of the nuclear SLR is more pronounced in underdoped cuprates where the difference
between $T^{\ast}$ where the pseudogap opens and $T_c$ can amount 100\,K and even more.  In
overdoped systems the $(T_1T)^{-1}$ product is relatively flat above $T^{\ast}$ and the discrepancy
between $T^{\ast}$ and $T_c$ is substantially diminished
\cite{ZhengPRL05,KitaokaPRB98,KitaokaPhC91}. As for the sample in study, the essential difference
between the temperature where the pseudogap opens and $T_c$ viewed in both the NMR line shift
(Fig.~\ref{Fig3}) and Cu SLR (Fig.~\ref{Fig4}) temperature dependencies, as well as the
hyperbolic-type $T$-dependence of the $(T_1T)^{-1}$ product above 60\,K, indicate a slightly
underdoped state of Bi$_{1.6}$Pb$_{0.4}$Sr$_{2.05}$CuO$_y$ system.

Summarizing, the Cu NMR data collected for Pb-doped Bi:2201 compound revealed the pseudogap that
opens at $T^{\ast}=20-60\,$K, way above the $T_c\simeq9\,$K measured for the same crystal
orientation and value of the NMR experiment field. The noticeable discrepancy between $T_c$ and
$T^{\ast}$ and the behavior of Cu SLR at $T>T^{\ast}$ imply the underdoped state of the studied
system. Quite unexpectedly, the magnetic field of 7\,T applied along the \textit{c} axis of the
crystal, shifted the zero-field $T_c=16\pm 1\,$K by only 7-8\,K. Usually, in other HTC cuprates
with similar zero-field $T_c$ the same magnetic field either completely destroys the
superconductivity or shifts the transition temperature below 4\,K \cite{GantJETPL00, ZverevJETP14}.
The relatively weak magnetic field effect on the superconductivity observed for
Bi$_{1.6}$Pb$_{0.4}$Sr$_{2.05}$CuO$_y$ indicates the unforeseen high value of the upper critical
field.

\section*{Acknowledgements}
The authors gratefully acknowledge preliminary ac-susceptibility measurements of the studied sample
performed by V.\,N.\,Zverev and A.\,V.\,Palnichenko. The work was partially supported by the
Russian Federal Program for Fundamental Research, Sec.II.8 ``Actual problems of Condensed Matter
Physics''.

\section*{References}

\bibliographystyle{model1-num-names}
\bibliography{BiPb2201_bibliography}

\end{document}